\documentclass[fleqn,usenatbib]{mnras}

\usepackage{newtxtext,newtxmath}

\usepackage[T1]{fontenc}

\DeclareRobustCommand{\VAN}[3]{#2}
\let\VANthebibliography\thebibliography
\def\thebibliography{\DeclareRobustCommand{\VAN}[3]{##3}\VANthebibliography}

\usepackage{graphicx}	
\usepackage{amsmath}	
\usepackage{hyperref}
\usepackage{xcolor}

\title[Relativistic Effects on Triple Black Holes]{Relativistic Effects on Triple Black Holes: Burrau's Problem Revisited}

\author[A. S. Chitan et al.]{
A. S. Chitan,$^{1}$\thanks{Email:ariel.chitan@my.uwi.edu}
A. Myll{\"a}ri,$^{2}$
and S. Haque$^{1}$
\\
$^{1}$Department of Physics, The University of the West Indies, Trinidad, W.I.\\
$^{2}$Department of Computers and Technology, St. George's University, Grenada, W.I.\\
}

\date{Accepted XXX. Received YYY; in original form ZZZ}

\pubyear{2021}

\begin{document}
\label{firstpage}
\pagerange{\pageref{firstpage}--\pageref{lastpage}}
\maketitle

\begin{abstract}

We explore, using numerical simulations, the influence of mass and distance on the evolution of triple black hole systems. Following in the direction of Burrau's famous 3,4,5 problem, black holes are initially placed at the vertices of Pythagorean triangles. Numerical integration of orbits was conducted using relativistic corrections (post-Newtonian) up to the 2.5$^{th}$ order with ARCcode. As a descriptor of the evolution of the systems, the lifetimes, the number of two-body  encounters and the number of mergers were all analysed. We found that as the mass unit of the black holes increased there was strong positive correlation with the fraction of mergers (0.9868), strong negative correlation with the average number of two-body encounters (-0.9016)  and the average lifetimes of the triple systems decayed exponentially (determination coefficient of 0.9986). Around the mass unit range of 10$^{5.5}$M$_{\odot}$-10$^{5.6}$M$_{\odot}$, there was a transition from escape dominated dynamics to merger dominated dynamics. However, in the mass unit range of 10$^6$ M$_{\odot}$ $-$ 10$^{9}$M$_{\odot}$ with 1 pc distance unit, we find that 25\% of cases resulted in the escape of a supermassive black hole (SMBH) which may be a cause for wandering SMBH's found in some galaxies/galactic merger remnants.

\end{abstract}
\color{black}



\section{Introduction}
It is thought that at the center of almost every massive galaxy there exists a supermassive black hole (SMBH) \citep{Kormendy1995}. Their much smaller counterparts – stellar black holes (SBH) are thought to be abundant throughout galaxies. Intermediate mass black holes (IMBH) have not been as easy to detect though they are postulated to exist. However, a very compelling candidate has been recently discovered for the existence of these types of black holes using data from the Hubble Space Telescope \citep{Lin2020}. The origin of IMBH’s is hypothesized to be of countless of interactions and mergers of smaller black holes in the early universe. And for the first time, LIGO/VIRGO systems have detected the merger of a binary pair of black holes (BBH) that resulted in an IMBH of 142 M$_{\odot}$, the confirmation of the existence (and formation) of a black hole within the IMBH mass range \citep{Abbott2020}. This is hypothesized to have lead to subsequent mergers and formation of SMBH’s \citep{Gebhardt2005}. How black holes interact and merge is important not just to understand the origin of SMBH’s but also due to advancements in gravitational wave detectors as the LIGO/VIRGO and upcoming LISA systems. LIGO and VIRGO detections have catapulted research in gravitational waves and black hole dynamics and the upcoming LISA has potential of detecting even fainter signals than LIGO/VIRGO. In a series of papers by \citet{Bonetti2018,Bonetti2019}, the merger rates for massive black holes were studied using 2.5th order post-Newtonian (pN) equations for hierarchical triple systems of massive black holes in the range of 10$^5$ M$_{\odot}$ $-$ 10$^{10}$ M$_{\odot}$ .  Bonetti et al. have shown that for the range of black holes simulated, up to seventy five merger events per year may be detectable by the LISA system with up to 30\% of these mergers having occurred due to the influence of a third black hole depending on the seeding model used. 

Triple systems of black holes are of particular interest as BBHs can be driven to merge on a much shorter timescale when there is the gravitational influence of a third, external black hole \citep{Fragione2019}. In hierarchical triple systems, when there is a distinct inner binary orbited by a distant third body, the eccentricity of the inner binary oscillates periodically due to the von Zeipel-Kozai-Lidov (ZKL) mechanism   (\citep{zeipel1909}, \citep{kozai1962}, \citep{lidov1962}). This allows for high eccentricities of the inner binary to be achieved which can then drive mergers of BBH's in these types of configurations. The effect of ZKL mechanism on small mass black holes in globular clusters was studied by \citet{miller2002} where formation of these hierarchical triple systems can lead to merger events instead of BBH's being kicked from the cluster due to other triple interactions. \citet{thompson2011} studied compact object binaries and showed that the timescales for mergers are drastically decreased if one considers that these binaries are actually part of hierarchical triples influenced by the ZKL mechanism. \citet{Blaes2002} also looked at the ZKL mechanism in SMBH triple systems. From their study they concluded that in bound hierarchical SMBH triples, that there was substantial decrease in merger time. In their dataset they found that merger time was reduced by a factor of ten in about 50\% of all cases for nearly equal mass hierarchical SMBH triple systems experiencing the ZKL mechanism. 

\color{black}

A third black hole can also aid in solving the final parsec problem, which describes how binary pairs of massive and supermassive black holes may never get close enough to merge \citep{MilosMerrit2003}. When galaxies merge, stellar and gaseous fields provide dynamical friction which slow the SMBHs enough so that they migrate to the center of mass. Triple interactions with stars allow the two SMBH’s to get close enough so that gravitational wave emission can take over and become the primary driving force for their merger \citep{Milo2003}. However, studies have found that at a certain distance, the number of stars available to drive the SMBH’s close enough are greatly diminished, due to ejections, and the two never get close enough to merge. The influence of a third black hole can replenish this loss of stars which would quicken the merger or it can directly influence the orbit of the BBH \citep{Ryu2017}.\color{black} Triple systems of SMBH's are expected to exist when three galaxies merge. A very recent discovery of such a system was found by \citet{Kollatschny2020}. NGC 6240, has been long known to harbour two active galactic nuclei but using a more detailed spectroscopic study with data from MERLIN and VLBA, three active galactic nuclei were resolved in the system. It was found that this system is in the final state of merging and all nuclei exist in a range of within 1 kpc with masses each in excess of \begin{math}9\times10^7 M_{\odot}\end{math}. 

Studying triple systems is more complicated than studying binary systems. As there exists no analytical solution to the three body problem, numerical orbital integration must be used to study individual cases. With modern technology, it has become possible to conduct numerical simulations and analyse the evolution of triple systems from a statistical standpoint. In 1913, Carl Burrau put forth his famous problem of three bodies. The evolution of three bodies was followed with initial conditions such that each of the bodies were placed at the vertices of a 3,4,5 Pythagorean triangle. The masses of each of the bodies also reflected the side of the triangle that was opposite to it  \citep{Burrau1913}. Without the use of computers, he was able to calculate, manually, the trajectory of these bodies but for a limited time. 

\begin{figure}
	\includegraphics[width=\columnwidth]{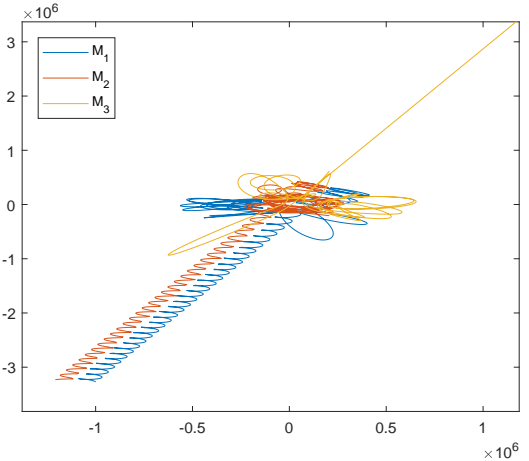}
    \caption{The orbital solution of Burrau's 3,4,5 three-body problem. Mass units are 3 M$_{\odot}$,4 M$_{\odot}$ and 5 M$_{\odot}$ with initial separations of the bodies being 3 pc ,4 pc and 5 pc (i.e. the distance unit is 1pc).}
    \label{fig:345orbital}
\end{figure}

Numerical research on the general three-body problem continued after this with notable works by \citet{Szebehely1967} who analysed Burrau’s problem with the use of computers. They found that the three body problem was not periodic nor quasi periodic but after looking at numerous cases, found that an eventual evolution of some three body systems resulted in a binary pair being formed and one of the other bodies escaping the system entirely. Hut and Bahcall, in a series of papers studied the gravitational scattering of binary-single star systems involving intensive simulations \citep{Hut1,Hut2,Hut3,Hut4,Hut5,Hut6,Hut7}. In the analytical treatment of three-body systems, a more statistical approach has demonstrated significant headway. This involves the work by \citet{monaghan1976} where the disruption of triple systems with low angular momentum was studied. A qualitative, statistical description of the escaping third body was obtained from this.
\color{black}
 \citet{valtonen1995} further studied Burrau’s problem in five different cases – varying the mass unit from 10$^5$ M$_{\odot}$ to 10$^9$ M$_{\odot}$. Their results showed that the smaller mass systems typically ended in a binary pair formation and the third body escaping the system. The large-mass cases ended in the eventual merging of all three bodies. In Figure \ref{fig:345orbital} we recreate the solution to Burrau's three-body problem where the bodies are of masses 3 M$_{\odot}$, 4 M$_{\odot}$ and 5 M$_{\odot}$ and start from rest at the vertices of a 3,4,5 Pythagorean triangle with a distance unit of 1pc. \color{black}In this solution, pN terms have been used up to the 2.5$^{th}$ order. \color{black} Bodies M$_1$ and M$_2$ form a binary pair and M$_3$ escapes. 

 Analytic research on the three-body problem has also been studied throughout the years alongside numerical research. Eccentricity changes in binaries due to an encounter with a third distant body have been studied analytically with particular applications to low-mass binary millisecond pulsars in globular clusters. These analytic expressions have been in great agreement with numerical results \citep{Heggie1996}. More recently, great advances have been made by \citet{stone2019}, where a statistical solution was found to the chaotic non-hierarchical three-body system using the ergodic hypothesis. This formalism derived will be of great use as they have found that with resonant encounters, their outcomes are in good agreement with numerical integrations. \citet{ginat2021} also recently presented an approximate, statistical solution for bound non-hierarchical triples using a random walk model which is in agreement with numerical simulations. Another recent study was conducted by \citet{Hamers2019} extending the work by \citet{Heggie1996}. They present analytic calculations focused on the secular effects of a binary on a pertuber body.
\color{black}
In the following paper we present the results of numerical simulations in the framework of Burrau's problem, as an extension of the study of \citet{valtonen1995}, where the mass unit range is now increased from 10$^0$ M$_{\odot}$ to 10$^{12}$ M$_{\odot}$. This paper is divided into 3 sections. \textbf{Section 2} will focus on the method and dataset. In \textbf{Section 3}, the results for the effect of relativistic corrections, the effect of mass unit and the effect of distance unit will be presented and the implications of these results will be discussed.

\section{Methodology and Dataset}

 Initial configurations for orbital integration are not only the 3,4,5 triangle but all Pythagorean triangles with c (hypotenuse) less than 100 making this study a total of sixteen triangles, each with thirteen individual cases. The code used was provided by Prof. Seppo Mikkola(\citep{Mikkola1993,Mikkola1996,Mikkola1999,Mikkola2002,Mikkola2006,Mikkola2008,Hellstrom2010,MikkolaElsev.2013,Mikkola2013}) which uses up to 2.5$^{th}$ order pN corrections for N-body simulations. Initially, higher order approximations were considered for this work, however, a recent study by \citet{Valtonen2017} has shown that higher order terms in post-Newtonian equations used for approximating triple body interactions may not be very accurate. In studying the BBH system, OJ 287,over a period of fourteen flares, an orbit search algorithm was used to map against recorded flares. It adjusted until a good model orbit could be found for the BBH.  It was found that for such an orbit, a solution only exists if the first radiation reaction term is used, the 2.5th order term, as higher order radiation reaction terms (3.5, 4.5) are close in absolute value but opposite in sign which result in cancellation.

We focus on three major parameters of the triple systems: (a) The number of mergers - It is most probable that triple systems can end in one of two ways: 1. Two of the bodies form a binary pair and the third escapes - we detect an escape as when a body obtains hyperbolic velocity with respect to the center of mass of the system; 2. All three bodies merge. A third, less typical outcome is that there is a merger of two bodies and the merger remnant and third body separate (this may be due to a relativistic kick after the merger of two black holes \citep{hughes2004}, \citep{baker2008}) Relativistic effects used in the code ensure that due to gravitational wave emissions, mergers will always ensue in place of bound binaries, though in some cases this may take longer than Hubble time. The binary pair formed in Figure \ref{fig:345orbital} will eventually end in a merger. ; (b) The lifetime of the triple systems- we define as the time until the first merger occurs or in the case where there are no mergers, the time until a permanent binary pair is formed and the third body escapes and ; (c) number of two-body  encounters among the three bodies, which we define as an (not always bound) binary interaction between any of the two bodies during the system's lifetime. 

\subsection{Timescales}

The crossing time of a system ($T_{CR}$) is defined as the time taken for a body to pass through a system as given by the equation below \citep{karttunen}

\begin{equation}
T_{CR}=GM^{5/2}(2|E_0|)^{-3/2}
\end{equation}

Here G is the gravitational constant, M is the total mass of the system and $E_0$, the energy of the system.
\color{black}

2000 crossing times were used as upper boundaries when running simulations instead of years since with the smaller mass systems, the upper limit in years was at times not long enough for interactions among the bodies within the system. For the data describing lifetimes of the systems, timescales were kept in terms of crossing time unless stated otherwise. 

\subsection{Two-Body  Encounters}

One of the parameters we consider is the number of two-body  encounters that occurs within a triple system's lifetime. This is employed by checking the minimum distances between bodies at each timestep. When a local minimum of this value is observed, we detect a two-body  encounter.

\begin{figure}
	\includegraphics[width=\columnwidth]{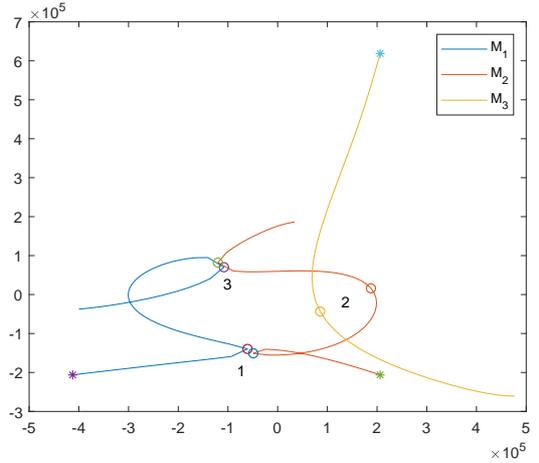}
    \caption{An illustration of three two-body  encounters labelled as 1,2 and 3. Two-body  encounters are discussed in Section 2.2.}
    \label{fig:BEIllustration}
\end{figure}

Figure \ref{fig:BEIllustration} shows the plot of a two-body  encounter. This is for the 3,4,5 case with mass unit 10$^{6}$ M$_{\odot}$ and distance unit 1pc. The plot illustrates the first 120 time steps within which three two-body  encounters are detected. The stars show the initial positions for the three bodies. The plot shows the bodies moving toward each other where M$_1$ and M$_2$ have a close encounter, labelled as 1, and subsequently move away from each other. This is the first two-body  encounter. The second and third are marked as 2 and 3 with an encounter between M$_2$ and M$_3$ and then again with M$_1$ and M$_2$, respectively.

We use the number of two-body  encounters as a parameter in order to characterise interaction within a system.

\color{black}
\subsection{Dataset}
 The effect of relativistic corrections was first briefly investigated using two triangles - (3,4,5) and (5,12,13) by comparing results with simulations done with relativistic corrections turned off. Simulations were conducted for each triangular configuration where mass was varied as 10$^0$ M$_{\odot}$ - 10$^{12}$ M$_{\odot}$. In the first set, relativistic corrections were turned on (up to pN 2.5$^{th}$ order) and the second set, with the same mass units, relativistic corrections were turned off for a purely Newtonian result. 
 
 Since Pythagorean triangles were used as startup configurations, initial separations of the bodies are always integer values in parsecs. 
 
 We then conducted in depth simulations on the classical 3,4,5 configuration where both mass and distance were varied. The mass unit was increased from 10$^0$ M$_{\odot}$ - 10$^{12}$ M$_{\odot}$ in increasing factors of lg(0.1 M$_{\odot}$). For each mass unit used, the distance unit was varied from 10$^{-1}$ pc to 10$^4$ pc in increasing factors of lg(0.5 pc).
 
Simulations were then run for a total of sixteen initial configurations (with relativistic corrections on): the sixteen Pythagorean triangles with c$<100$. For each triangle, thirteen individual simulations were run so that the mass unit could be varied for each from 10$^0$ M$_{\odot}$ - 10$^{12}$ M$_{\odot}$, the distance unit was kept at 1 pc. A summary is provided in Table \ref{tab:summary}. 

Parameters used to compare each of the systems are: 1. The fraction of mergers and whether a system moves from being three bodies to being two via either a merger or an escape, 2. The lifetimes of the systems, 3. The number of two-body  encounters. In all simulations conducted, point masses all began with zero initial velocity, zero angular momentum and were kept in the 2D planar case.

\begin{table*}
 \caption{A summary of simulations conducted separated into individual Runs.}
\begin{center}
 \begin{tabular}{ ccccc } 
 \hline
 Run & Influence of & Configuration Tested & Mass Unit (M$_{\odot}$) & Distance Unit (parsecs) \\ [0.5ex] 
 \hline
 One & Relativistic Effects & (3,4,5), (5,12,13) & 10$^0$, 10$^1$,..., 10$^{11}$, 10$^{12}$ & 1 \\ 
 \hline
 Two & Mass and Distance & (3,4,5) & 10$^0$ , 10$^{0.1}$, ..., 10$^{11.9}$, 10$^{12}$ & 10$^{-1}$, 10$^{-0.5}$,..., 10$^{3.5}$, 10$^{4}$ \\
 \hline
 Three & Mass & All Pythagorean Triangles with Hypotenuse$<$100 & 10$^0$, 10$^1$,..., 10$^{11}$, 10$^{12}$ & 1\\
 \hline
 Four & Distance & All Pythagorean Triangles with Hypotenuse$<$100 & 10$^6$ & 10$^{-2}$, 10$^{-1}$,..., 10$^{3}$, 10$^{4}$ \\
 \hline
\end{tabular}
\end{center}
\label{tab:summary}
\end{table*}
\color{black}
\section{Results and Discussion}
\subsection{Effect of Relativistic Corrections}
\begin{figure}
	\includegraphics[width=\columnwidth]{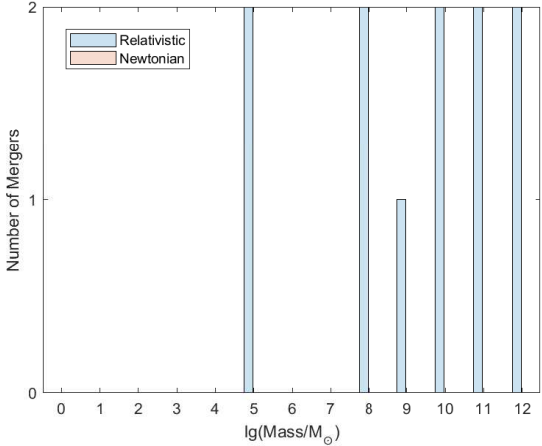}
    \caption{The number of mergers per each mass unit for the 3,4,5 triangle in the Newtonian and Relativistic cases. In the Newtonian case there are no mergers.}
    \label{fig:NewtMerge345}
\end{figure}

\begin{figure}
	\includegraphics[width=\columnwidth]{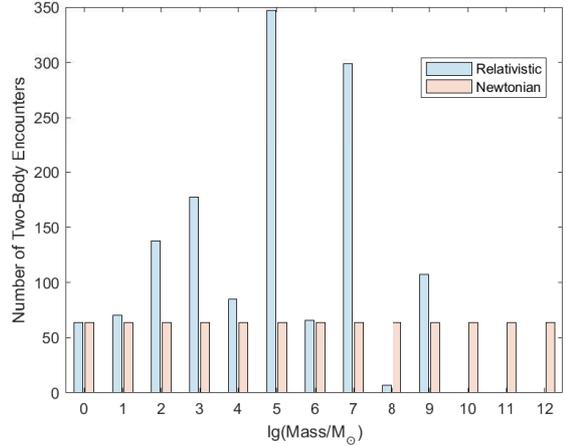}
    \caption{The number of two-body  encounters per each mass unit for the 3,4,5 triangle in the Newtonian and Relativistic cases.}
    \label{fig:NewtBE345}
\end{figure}

In order to verify the effect of relativistic corrections on the evolution of these systems we simulate first without relativistic corrections turned on, for a Newtonian perspective, and then with corrections turned on, for a relativistic perspective for the (3,4,5) triangle. Figure \ref{fig:NewtMerge345} shows the number of mergers for both cases - no merging occurs in the Newtonian simulations but merging appears from the IMBH range in the relativistic simulations. Figure \ref{fig:NewtBE345} shows the number of two-body  encounters in the Newtonian case as compared to the very interactive relativistic one. Obviously, Newtonian cases are not mass dependent as the relativistic cases.

\begin{figure}
	\includegraphics[width=\columnwidth]{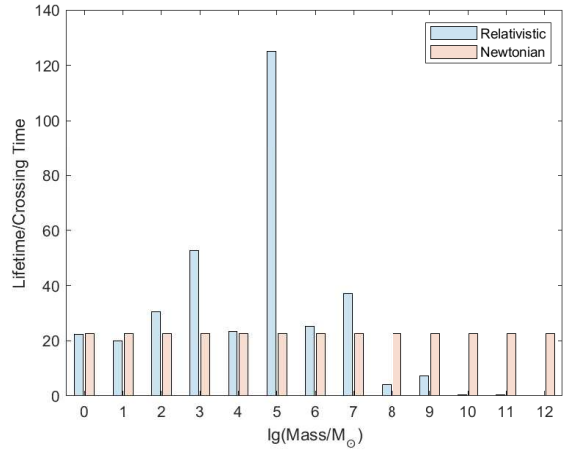}
    \caption{The lifetime per each mass unit for the 3,4,5 triangle in the Newtonian and Relativistic cases.}
    \label{fig:NewtLifetime345}
\end{figure}
There are obvious spikes at the 10$^5$ M$_{\odot}$ and 10$^7$ M$_{\odot}$ mass units. These systems experience several long ejections and binary pair formations which ensures longer lifetimes and more two-body  encounters detected than the other systems. The spike at 10$^9$ M$_{\odot}$ occurs due to the highly chaotic nature of this system. As compared to the preceding (10$^8$ M$_{\odot}$) and following (10$^{10}$ M$_{\odot}$) systems, this one does not experience merging of all three bodies within 2000 crossing times. Instead there is potentially a relativistic kick as two of the bodies (M$_2$ and M$_3$) merge which then allows the third body (M$_1$) and the remnant to go separate ways shortly after. In the 10$^8$ M$_{\odot}$ and 10$^{10}$ M$_{\odot}$ mass units two mergers occur in much shorter times. The typical prediction would be that at the 10$^9$ M$_{\odot}$ mass unit there would also be two mergers, however due chaotic nature of the three-body system, this does not occur. 

Figure \ref{fig:NewtLifetime345} shows the lifetime (crossing times) in the relativistic approximation vs. Newtonian. As expected, the Newtonian cases do not depend on the increase in mass and changing crossing time scale. This is not reflected in the relativistic cases which show a marked dependence on mass.

 The influence of mass increase is seen particularly in the dominance of merging. Relativistic effects - namely the emission of gravitational radiation introduced at the 2.5th pN order - make the systems merge very quickly at larger mass units and this makes the evolution of such systems much more regular and less complicated than in the Newtonian cases. The results using the (5,12,13) configuration exhibit similar results.

\subsection{The 3,4,5 Triangle}
\begin{figure}
	\includegraphics[width=\columnwidth]{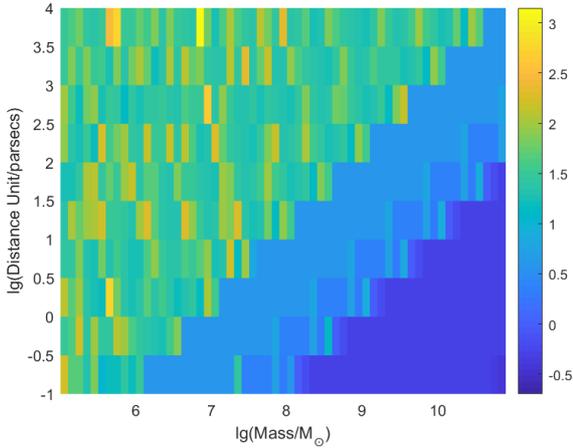}
    \caption{The lg(Lifetime/crossing time) for 3,4,5 configuration as mass unit and distance unit are increased for mass units 10$^5$ M$_{\odot}$ to 10$^{10.9}$ M$_{\odot}$.}
    \label{fig:lg(life)_all}
\end{figure}

Before studying the ensemble of Pythagorean triples we look at the 3,4,5 triangle in greater detail than had been done by Valtonen et al. The lifetimes of the systems as mass and distance are varied are shown in Figure \ref{fig:lg(life)_all}.
The mass units 10$^5$ M$_{\odot}$ to 10$^{10.9}$ M$_{\odot}$ were chosen to show the distinction between smaller mass systems and large-mass systems. Because some mass units higher than 10$^{10.9}$ M$_{\odot}$ had instantaneous mergers and ended their lifetimes immediately, the choice of logarithmic scaling prevented them from being presented. Using the logarithmic scaling of the lifetimes, however, allows us to present data without modifying/removing outliers. It can be seen that for masses 10$^5$ M$_{\odot}$ and 10$^6$ M$_{\odot}$, for all distance units the lifetimes are not only longer than mass units onward but also much more variant and unpredictable. The mass units between 10$^6$ M$_{\odot}$ and 10$^{10.9}$ M$_{\odot}$ show a starker dependence on the distance unit. Large mass systems, for example the 10$^9$ M$_{\odot}$ unit, at shorter distance units (from 10$^{-1}$ pc to 10$^2$ pc) have predictable and shorter lifetimes, but as the distance unit is increased, a similar behaviour to the smaller mass systems in terms of length and unpredictability of lifetime begins. This is a trend that can be expected of even the largest mass units studied here if the distance units are increased further than 10$^4$ pc. Though it should be noted that triple systems of SMBH's $>$ 10$^{10}$ M$_{\odot}$ forming on sub-parsec distance scales are more of an academic interest as they may be unlikely in nature.

\begin{figure}
	\includegraphics[width=\columnwidth]{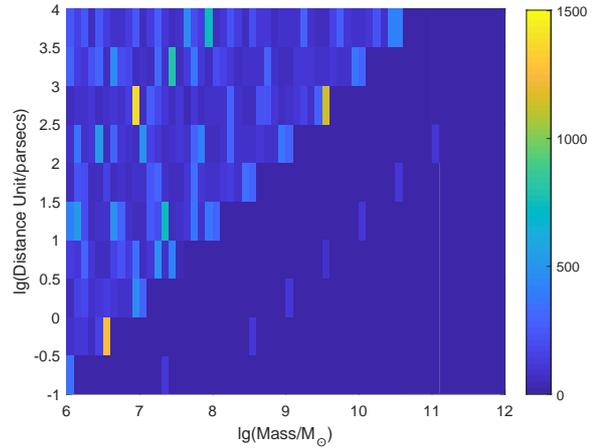}
    \caption{The number of two-body  encounters for 3,4,5 configuration as mass unit and distance unit are increased for mass units 10$^6$ M$_{\odot}$ to 10$^{12}$ M$_{\odot}$.}
    \label{fig:BE6_12}
\end{figure}

This trend is also shown in the number of two-body  encounters as in Figure \ref{fig:BE6_12}. For presentation of this graph, one outlier (values outside of 3  standard deviation's from the mean within each mass range) had to be removed. This configuration had mass unit 10$^{6.8}$ M$_{\odot}$ and distance unit 10$^{3.5}$ pc with 5403 two-body  encounters were detected. Smaller mass systems exhibit much more interaction than the large-mass systems which, due to the greater emission of gravitational radiation, are attracted to each other much more quickly and prefer merging over interaction. The three yellow peaks in the graph show configurations of high numbers of detected two-body  encounters. These systems are 1. mass unit 10$^{6.5}$ M$_{\odot}$ and distance unit 10$^{-0.5}$ pc, 2. mass unit 10$^{6.9}$ M$_{\odot}$ and distance unit 10$^{2.5}$ pc and 3. mass unit 10$^{9.5}$ M$_{\odot}$ and distance unit 10$^{2.5}$ pc. These systems, along with the outlier mentioned above, experience many binary interactions between two of the bodies while the third body has a very long ejection. As the third body is ejected for a long duration, it returns slowly to the system where another binary pair formation happens after much more interplay. This explains why these systems show such high numbers of encounters.

\begin{figure}
	\includegraphics[width=\columnwidth]{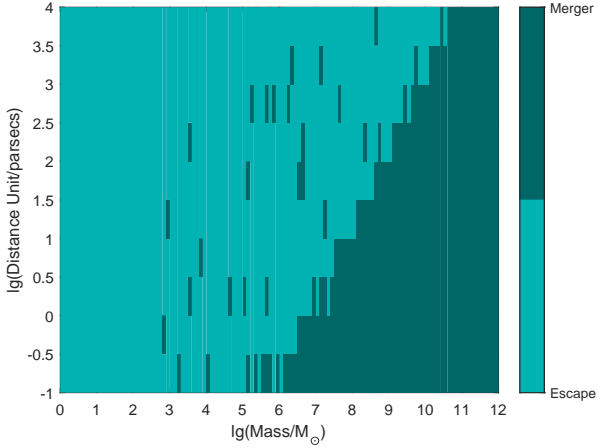}
    \caption{The simulations which end in escape or merger for the 3,4,5 configuration.}
    \label{fig:345EscapeVMerger}
\end{figure}

We are also interested in the manner in which the systems end their lifetimes - if this is an escape or a merger. From Figure \ref{fig:345EscapeVMerger}, as had been found by \citet{valtonen1995}, the smaller mass systems end in an escape of one of the bodies while the large-mass systems end in a merger of two of the bodies. This is expected since the smaller mass systems behave in a more Newtonian manner than the large mass systems. However, the chaotic nature of the 3,4,5 triple system manifests itself as even at the small mass units there are systems which end in mergers. These outcomes seem almost random and demonstrate how changing the initial conditions very slightly can produce different events. For example, the 10$^{2.9}$ M$_{\odot}$ system at the 10$^1$ pc distance unit does not experience a merger nor an escape during the 2000 crossing times the simulation was run for. This system is outstanding since its lifetime is the  greatest. The third body experiences a very long ejection for the entire simulation. In Figure \ref{fig:345EscapeVMerger}, this system is listed as escape solely for the construction of the figure. The 10$^{2.8}$ M$_{\odot}$ at 10$^{0.5}$pc is the only system of this mass unit that ends in a merger. The 10$^{3.5}$ M$_{\odot}$ system at 10$^0$pc and 10$^2$pc ends its triple lifetime in a merger, however, this is strange since even when this system starts off with a sub-parsec distance there is escape over merger. This system (and the others that also ended in unexpected ways) also showed similar trends in their lifetimes and the number of two-body  encounters they had. 
\color{black}

If we consider Figures \ref{fig:lg(life)_all}, \ref{fig:BE6_12} and \ref{fig:345EscapeVMerger}, The clear diagonal lines may be derived by comparing the crossing time with the gravitational wave merger time.

From \citet{Peters1964}, the gravitational wave merger time is
 \begin{equation}
T_{GWmerge}=\frac{3\alpha}{85}\left(\frac{1}{G^3m_1m_2(m_1+m_2)}\right)
\end{equation}

where $\alpha$ is
\begin{equation}
    \alpha=(a_0^4c^5)(1-e_0^2)^{7/2}
\end{equation}

Here, $m_1$ and $m_2$ are the merging bodies, $a_0$ and $e_0$ are the initial semi-major axis and initial eccentricity of the two bodies respectively and c is the speed of light. The crossing time can be defined as:
\begin{equation} \label{eq:crt2}
    T_{CR}=\frac{R}{V}
\end{equation}

Where V is the velocity of a body passing through the system of dimension R \citep{karttunen}. For the uniform sphere of total mass M (here the mass of the three bodies) and radius R, V becomes
\begin{equation}
    V=\sqrt{\frac{GM}{R}}
\end{equation}

So, the crossing time can be written as:
\begin{equation}
    T_{CR}=\frac{R^{3/2}}{\sqrt{GM}}
\end{equation}

Equating the crossing time with the gravitational wave merger time yields:
\begin{equation}
    lg(R)=\frac{1}{3}lg(M)+\frac{1}{3}lg\left(\frac{k^2}{G^{5}}\right)
\end{equation}
where 
\begin{equation}
    k=\frac{3\alpha}{85m_1m_2(m_1+m_2)}
\end{equation}

So, we obtain a linear relationship between lg(R) and lg(M). This can approximately describe the slopes observed as in  Figures \ref{fig:lg(life)_all}, \ref{fig:BE6_12} and \ref{fig:345EscapeVMerger}, the lg(Distance Unit) and lg(Mass Unit) are plotted. 
\color{black}
\subsection{Effect of Increasing Mass} \label{Effect of Increasing Mass}
\begin{figure}
	\includegraphics[width=\columnwidth]{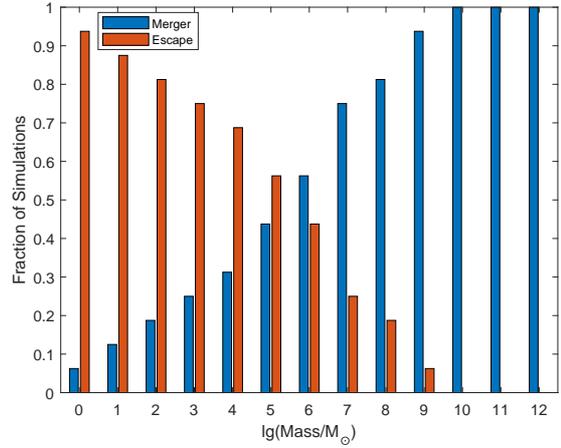}
    \caption{The fraction of simulations which end in a merger or an escape.}
    \label{fig:mergeorescape}
\end{figure}

For each mass unit studied there are 32 mergers possible. There are 16 individual simulations per mass unit and in each simulation, two mergers in total are possible within 2000 crossing times. If we consider what fraction of these possible 32 mergers take place within the 2000 crossing times as the mass unit increases, a correlation coefficient of 0.9868 exists between the mass unit and the fraction of mergers suggesting strong, positive correlation between the two variables. A notable increase is seen from the 10$^7$M$_{\odot}$ mark which also appears in Figure \ref{fig:mergeorescape} regarding the fraction of simulations that end in mergers (blue bars). This is in corroboration with the results of \citet{valtonen1995} where mergers begin to dominate from about the same mass unit. It is here that the problem transitions from a Newtonian scenario to a relativistic one. Gravitational radiation plays a bigger role and the bodies lose energy more quickly and subsequently fall inward and merge at a much faster rate.

\begin{figure}
	\includegraphics[width=\columnwidth]{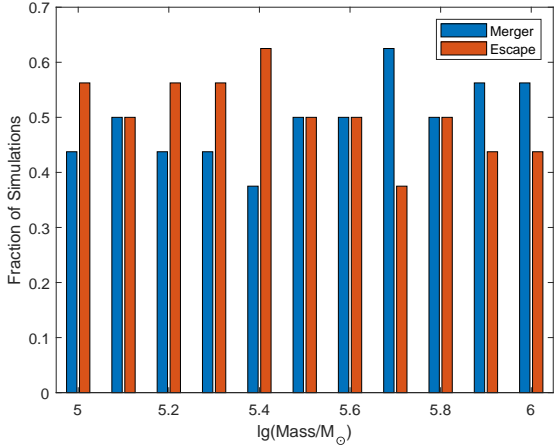}
    \caption{The fraction of simulations which end in a merger or an escape between 10$^{5}$M$_{\odot}$-10$^{6}$M$_{\odot}$.}
    \label{fig:mergeorescapelg5tolg6}
\end{figure}

\begin{figure}
	\includegraphics[width=\columnwidth]{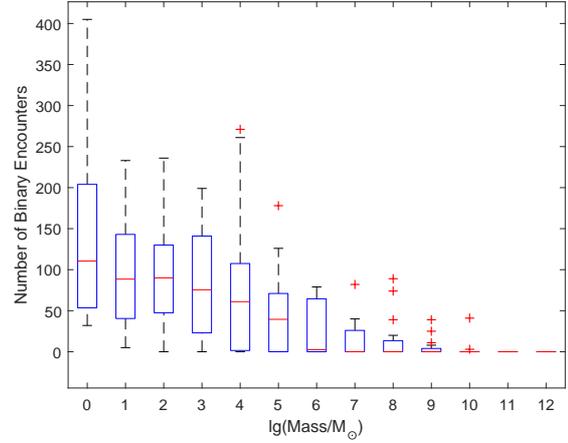}
    \caption{Box plot of the average number of two-body  encounters per each mass unit.}
    \label{fig:BE_box}
\end{figure}

We also consider the fraction of simulations which end their lifetimes in either a merger or an escape. This is seen in Figure \ref{fig:mergeorescape}. A high positive correlation between mass unit and the fraction of simulations which end in a merger was found (0.9838) and, correspondingly, a high negative correlation between the mass unit and fraction of simulations ending in an escape was found (-0.9838). A transitory mass range can be seen from escape to merger domination at 10$^5$M$_{\odot}$ - 10$^6$M$_{\odot}$. We look at this range more in depth by conducting simulations with increasing mass unit (10$^{5}$,10$^{5.1}$,..., 10$^{5.9}$,10$^{6}$)M$_{\odot}$ for all triangular configurations and find that the transition can be further narrowed down to the 10$^{5.5}$M$_{\odot}$-10$^{5.6}$M$_{\odot}$ range (Figure \ref{fig:mergeorescapelg5tolg6}). 
\color{black}
It was found that as the mass unit increases the average number of two-body  encounters per mass unit tend to decrease. A correlation coefficient between the two variables is -0.9016. In the box and whisker plot of Figure \ref{fig:BE_box}, for each mass unit, the central red line indicates the median value while the top and bottom edges of the box represent the 75$^{th}$ and 25$^{th}$ percentiles. The whiskers show the furthest datapoints not considered outliers while the red crosses depict outliers. \color{black} However, this plot was done after removing outliers (within each mass unit grouping) and replacing with the average number of two-body  encounters per mass unit. This was done solely for the legibility of the plot - otherwise the scaling made the values for the larger mass units impossible to see. These systems are listed in Table \ref{tab:outliers}. The larger the bodies become and the stronger the gravitational effect of each, the less there is interaction.  (including outliers omitted in Figure \ref{fig:BE_box}). These bodies, as seen by the merge data, tend to gravitate toward each other very quickly (on the largest scales, this is almost immediate) and coalesce.  This in turn affects the lifetimes of the systems and it is found then that the large-mass systems end their lifespans quickly while small systems last longer and the bodies interact with each other more. From the box and whisker plot of Figure \ref{fig:BE_box}, smaller mass systems show more variation in the number of two-body  encounters while the number of two-body  encounters for larger mass cases are more in agreement with each other. 
\begin{table}
 \caption{A list of outliers for the boxplot of Figure \ref{fig:BE_box}}
\begin{center}
 \begin{tabular}{ ccc } 
 \hline
 Configuration & Mass Unit/M$_{\odot}$ & Number of Two-Body  Encounters\\
 \hline
 (3,4,5)&10$^{5}$, 10$^{7}$, 10$^{9}$&348, 300, 108\\
 \hline
 (5,12,13)&10$^{4}$&537\\
 \hline
 (8,15,17)& 10$^{3}$& 689\\
 \hline
 (9,40,41)&10$^{2}$&460\\
 \hline
 (11,60,61)&10$^{1}$&564\\
 \hline
 (12,35,37)&10$^{3}$& 1512\\
 \hline
 (13,84,85)&10$^{0}$&1813\\
 \hline
  (28,45,53)&10$^{4}$&530\\
 \hline
 (33,56,65)&10$^{6}$, 10$^{7}$&232, 174\\
 \hline
 (48,55,73)&10$^{6}$&456\\
 \hline
 \end{tabular}
 \end{center}
 \label{tab:outliers}
 \end{table}

\begin{figure}
	\includegraphics[width=\columnwidth]{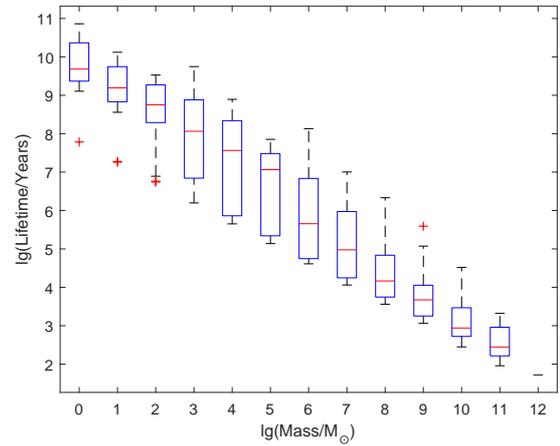}
    \caption{Box plot of the lg(lifetime/years) as mass unit is increased.}
    \label{fig:boxlife}
\end{figure}

We also consider the variation in lifetime for the ensemble of triples as in Figure \ref{fig:boxlife}. The symbol description is the same as in Figure \ref{fig:BE_box}. In both the small mass cases and large mass cases there is a general closeness in the length of lifetimes. As masses get intermediate, however, the variations appear to be more significant. The outliers, however, appear more at the small mass cases - a characteristic of the more chaotic small mass systems. The large mass systems, as already discussed, merge quickly and behave more expectedly. By curve fitting the data for average lifetimes (in years) of all the systems vs. mass unit, the average lifetime decays exponentially as \begin{math}(1.567\times10^{10})e^{-1.449x} \end{math}with a coefficient of determination of 0.9986.

If we consider the triangles individually, for comparative purposes, we find that with the exception of (9,40,41), (11,60,61), (16,63,65) and (13,84,85) triangular configurations, mergers do not occur for the smaller mass units i.e. 10$^0$M$_{\odot}$ to about 10$^3$M$_{\odot}$. The mentioned configurations that do not obey this all have one dimension that is significantly smaller than the other two. 

These would be considered closer to hierarchical type systems. Here, we follow the terminology of \citet{Anosova1990}.These systems though not hierarchical by definition, are still much closer to this type of configuration than the other Pythagorean triples discussed below and we use the description 'closer to hierarchical' to highlight the contrast in results. These systems form temporary binaries in the beginning of their lifetimes and describing systems in such a way has been done before \citep{Chernin1998}. There is obvious hierarchy though this is not the standard definition. Their dynamics and final evolutionary paths follow more closely to a traditional hierarchical system and as such we use this to separate the systems based on their behaviour. Binary pairs within typical hierarchical triples merge in a shorter time than binaries in non-hierarchical triples. In the cases listed above, merging dominates from even small mass units with the (11,60,61) triangle having mergers from even the 10$^1$M$_{\odot}$ mass unit.

In order to quantify in an informal manner which of the triangles are closer to hierarchical than others, the angle between the hypotenuse and the second longest side of the triangle can be used. For an angle that is $>$30$^{\circ}$, the triangle can be considered more non-hierarchical. This corresponds to the following triangles: (3,4,5); (20,21,29); (28,45,53); (33,56,65); (48,55,73); (65,72,97). Three of these triangles behave typically - (28,45,53), (33,56,65) and (65,72,97) - the two former triangles however, lie on the cusp of what we consider here to be hierarchical (31.9$^{\circ}$ and 30.5$^{\circ}$ respectively). The (65,72,97) triangle however has an angle of 42$^{\circ}$. It is of note that the (3,4,5) and the (48,55,73) triangles exhibit the strangest results, especially when looking at merge rates. 
The (20,21,29) triangle also behaves differently as it is the configuration that takes the longest for mergers to show up - this is at 10$^{10}$M$_{\odot}$. 

 Non-hierarchical triples can be more unpredictable than hierarchical triples. It has been found by extensive study (in the Newtonian case) that while there is some stability for true hierarchical systems, for non-hierarchical ones no criterion of stability exists \citep{Anosova1992}. Typically we can see that the more hierarchical triples (such as (11,60,61), (9,40,41), etc) end quickly with mergers dominating from even smaller mass units. 
 
 This is not so for the more non-hierarchical triples (such as the (20,21,29), (65,72,97), etc) where mergers only dominate at large mass units and lifetimes are longer. The (3,4,5) triangle demonstrates this pattern. This relationship is easily seen from the more hierarchical (11,60,61) triangle where mergers dominate from the 10$^1$M$_{\odot}$ and two-body  encounters decline at the same mass unit.

To predict how the non-hierarchical triples would evolve is difficult as it is a natural feature of the three body problem to exhibit chaos and unpredictability. \citet{PortegiesZwart2018} found that the 3,4,5 Pythagorean problem was time reversible from a Newtonian perspective (this, however, is not prevented by chaos).  It was also found by \citet{Boekholt2020} that five percent of triples of black holes with masses of 10$^6$M$_{\odot}$ and separations of 1pc and zero angular momentum) are fundamentally unpredictable due to being time irreversible up to Planck length in the Newtonian case.

The (3,4,5) simulations (distance unit kept at 1pc) when compared to that of Valtonen et al.(1995), show peculiar differences in the 10$^5$M$_{\odot}$ case. In our study, as mass increases the number of mergers dominate from 10$^8$M$_{\odot}$ but at the 10$^5$M$_{\odot}$ mass unit, there are two mergers. In the 1995 study, this was not the case. Mergers dominated from the \begin{math}8\times\end{math}10$^6$M$_{\odot}$  while escapes dominated at smaller mass units. At the 10$^9$M$_{\odot}$ mass unit, merging of all bodies occurred very quickly. In this study we see that only one merger occurred in 2000 crossing times for 10$^9$M$_{\odot}$ case. These differences can be explained by the updated code that was used between then and now by Prof. Mikkola. Updates would have been from 1995 to present time, 2020 (\citep{Mikkola1993,Mikkola1996,Mikkola1999,Mikkola2002,Mikkola2006,Mikkola2008,Hellstrom2010,MikkolaElsev.2013,Mikkola2013}).  The code used in 1995 by Valtonen et al. used a different method for orbital integration - a chain regularization method where Kustaanheimo-Stiefel co-ordinate transformations were used for regularization \citep{Mikkola1993}. The code used in this work, following several key updates, uses a logarithmic-Hamiltonian leapfrog algorithm instead that is much simpler and more accurate \citep{Mikkola2013}. These changes, due to the chaotic nature of the problem, could have been amplified to result in different outcomes.

In particular the 10$^5$ M$_{\odot}$ mass unit system is interesting as it fails numerical accuracy tests when the accuracy parameter, $\epsilon$, is varied (halved and doubled). In both cases when this value is changed, the system ends similarly to the 1995 solution where there is escape and binary pair formation instead of mergers. There is a noticeable divergence in trajectory around 32-51 crossing times between the simulation using standard accuracy parameter and those when the value is doubled and halved. During this period in the standard simulation there is a triple interaction which does not occur in the test simulations. $M_3$ is intercepted by $M_2$ while $M_1$ is nearby. This ultimately results in $M_2$ and $M_3$ merging which does not happen in the test simulations.  

This illustrates the fundamental unpredictability of this system. This system may fall into the fraction of simulations that are time irreversible in the Newtonian case. The mass unit 10$^5$M$_{\odot}$ is close to that of which was simulated in the work of \citep{Boekholt2020}, 10$^6$M$_{\odot}$. By changing the accuracy parameter (as had been demonstrated in that study) here, in our simulations, this results in unpredictability of the outcome.

\subsection{Effect of Distance}
\begin{figure}
	\includegraphics[width=\columnwidth]{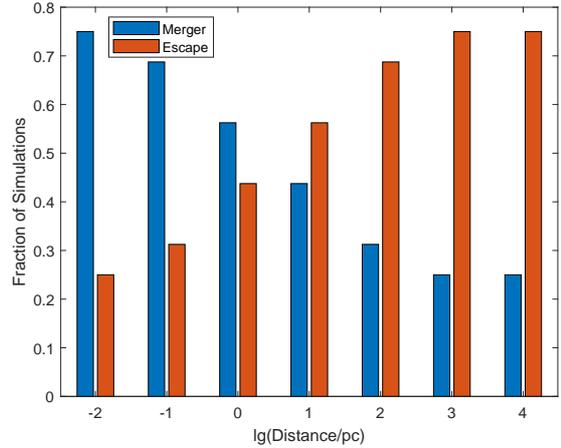}
    \caption{The fraction of simulations which end in and escape or a merger as distance unit is increased.}
    \label{fig:mergerescapedistance}
\end{figure}

The distance unit also plays a role in the evolution of the triple systems as was seen in the 3,4,5 configuration. Here we study how this affects the ensemble of triple configurations. There is strong negative correlation (-0.9791) between the fraction of simulations which end in mergers and increasing distance unit and, correspondingly, strong positive correlation (0.9791) between the fraction of simulations which end in escape and increasing distance unit with the mass unit fixed at 10$^6$M$_{\odot}$. For this mass unit we see that at the 10$^0$pc to 10$^1$pc distance units there is a transition from simulations ending in mergers to simulations ending in escape (Figure \ref{fig:mergerescapedistance}).
There is an exponential increase in average lifetime as the distance unit is increased at mass unit 10$^6$M$_{\odot}$ with equation \begin{math}(1.828\times10^{8})e^{2.776x} \end{math} with a coefficient of determination of 1.

\begin{figure}
	\includegraphics[width=\columnwidth]{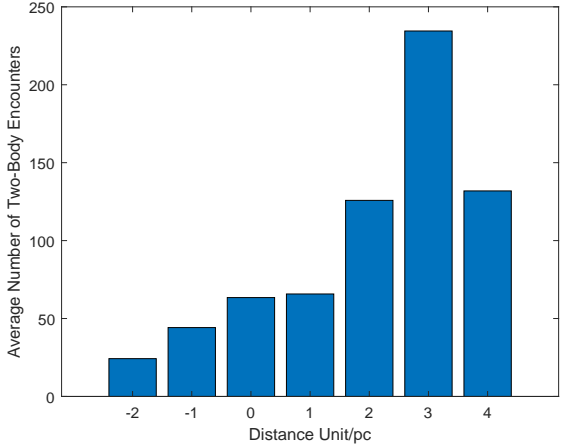}
    \caption{The average number of two-body  encounters as distance is increased}
    \label{fig:BEdistance}
\end{figure}

We also find that there is a spike in the average number of two-body  encounters at the 10$^3$pc distance unit for the mass unit studied. This may be the optimum distance unit (of all the units studied here) for the mass unit of 10$^6$M$_{\odot}$ to have the most interactive triple lifetime. The distance may be short enough that there is strong gravitational interaction but not too short that all of the black holes are immediately influenced to merge. This is shown in Figure \ref{fig:BEdistance}.

\subsection{Astrophysical Implications}
Below we include a brief discussion on the possible astrophysical implications of these findings. We focus mainly on the findings and the dataset of Section \ref{Effect of Increasing Mass}.

\subsubsection{SMBH triples}
SMBH triples are thought to form after the merger of three galaxies. When two galaxies merge, their respective SMBH's in the nuclei eventually form a binary pair. If a third galaxy merges with this merger remnant then a third SMBH is introduced and a triple SMBH system can form \citep{hoffman2007}.  We found strong positive correlation between the increasing mass unit and the number of simulations ending in merger over escape. SMBH's of the mass unit range 10$^{10}$M$_{\odot}$ - 10$^{12}$M$_{\odot}$ always resulted in mergers over escapes even at the largest distance unit under study (4pc). We expand on the cases of SMBH's $>$ 10$^{6}$M$_{\odot}$, both escape and merger scenarios, below.

SMBH's in the range of 10$^{6}$M$_{\odot}$ - 10$^{9}$M$_{\odot}$ while for the majority of simulations end in mergers, there are a few cases where one of the SMBH's can escape from the triple system (See Figure \ref{fig:mergeorescape}). In this mass unit range, $~$25\% of simulations resulted in escape of one of the SMBH's. One of these escapes was due to a potential relativistic kick post merger while the rest occurred in typical three-body fashion with an escaper and a binary pair (which would eventually end up merging).

The few triple SMBH systems that end in escape are particularly interesting as these types of systems can result in the so called 'wandering' SMBH. \citet{Reines2020} looked at dwarf galaxies and found massive black holes wandering around the outskirts of these galaxies. They found that in dwarf galaxies, these massive black holes are not necessarily hosted at the nuclei. This could be potentially due to a triple interaction in the past with black holes within these dwarf galaxies. \citet{Pesce2021} also found their so named 'Restless Supermassive Black Hole' (J0437+2456) of around 10$^{6}$M$_{\odot}$ wandering around its own galaxy. These types of wandering black holes could have escaped from a triple system at the center of the galaxy with great velocity uncharacteristic of typical SMBH's. It could also be that this wandering SMBH received a relativistic kick after undergoing a merger. The galaxy CXOC J100043.1+020637, also known as CID-42, also seems to be a merger remnant with a potentially recoiling SMBH \citep{Comerford2009}. One of the possibilities for this galaxy is that the AGN detected is a SMBH that either escaped from a triple system at the nucelus of the galaxy or received a relativistic kick after two other SMBH's merged at the center. Another possible scenario is that it is part of a recoiling binary \citep{Blecha2012,Civano2012}. These types of wandering black holes, per this study, should not be a rare occurrence as we know that triple SMBH systems do form in nature following galactic mergers.

To see whether or not these escapee SMBH's can be kicked from their host galaxies entirely would require further simulations involving stellar and other galactic medium along with the triple. In the simulations presented here, all are isolated triples. This is unlike the realistic scenario where external bodies may perturb the triple. However, other studies involving these types of simulations have found that the relativistic kicks that these SMBH's may receive post merger can be $~$4000-5000km/s which  may be enough to kick these SMBH's from their host galaxies entirely \citep{Campanelli2007,Gonzalez2007,Baker2007}.

Systems in the mass unit range of 10$^{6}$M$_{\odot}$ - 10$^{12}$M$_{\odot}$, mostly end in mergers of all three SMBH's. In these simulations the number of two-body encounters before mergers is much less for black holes of this magnitude, suggesting that not much interaction is needed to drive mergers due to the already strong gravitational effects. In fact, as the SMBH's become even larger, these effects become so strong that there is no real interplay among the black holes and they simply sink close to one another and merge. Such systems may always end up merging. In this case there should be a significant population of merging SMBH's that can be part of triple systems which we may be able to detect easily one day with sensitive gravitational wave detectors like LISA.

\subsubsection{SBH and IMBH triples}

Globular clusters are extremely active regions for stars, SBH's and potentially IMBH's. The first, concrete detection of an IMBH has been from the LIGO/VIRGO systems - event GW190521 \citep{Abbott2020}. Globular clusters, then, may harbour many triple systems (bound and unbound) and be great sources for gravitational wave signatures. \citet{Fragione2019} have presented a study on triple systems involving IMBH's and SBH binaries in hierarchical triples. They found that triple induced mergers between SBH's due to the induced ZKL mechanism should not be a rarity in globular clusters. Typically, these types of mergers would be detectable by LIGO/VIRGO systems.

In this study, SBH's and IMBH's typically prefer escape over merger. Though it is expected that after an escape the remaining two black holes will merge - though this may take longer than even Hubble time. These systems are also very long lasting and very interactive in comparison to the larger cases. In the mass unit range of 10$^{0}$M$_{\odot}$ - 10$^{5}$M$_{\odot}$, $~$23\% of simulations end via a merger (See Figure \ref{fig:mergeorescape}). This is a substantial percentage and suggests that triple systems of SBH's and IMBH's can produce merger events involving all three of the black holes. These types of triple systems are not rare as we expect their formation in globular clusters. Thus, mergers that we detect from our current gravitational wave detectors of black holes of these sizes may potentially be part of active triple systems which may result in two consecutive mergers. In these scenarios, IMBH's are left as remnants. 

Of worthy mention are also specific triangular configurations like (9,40,41), (11,60,61), (16,63,65) and (13,84,85). We found that with these triangular configurations, mergers occurred for the smaller mass units i.e. 10$^0$M$_{\odot}$ to about 10$^3$M$_{\odot}$. As previously discussed, these systems are not technically hierarchical but close to. Their outcomes are also suggestive of hierarchical type systems. It may be that when SBH and IMBH triples form in these types of configurations there is a more likely chance for merger. Of the 23\% of simulations mentioned, which ended in merger over escape, 95\% of these occurred in the more hierarchical configurations.
\color{black}
\section*{Conclusions}
By conducting simulations of triple black hole systems, it was found that as the mass unit increases, the problem becomes more relativistic around the 10$^7$M$_{\odot}$ unit and the number of mergers and the merger rate increase. High positive correlation was found between the fraction of mergers occurring and the mass unit (0.9868). This is opposite for the average number of two-body  encounters where high negative correlation was found with increasing mass unit (-0.9016). Around the mass unit range of 10$^{5.5}$M$_{\odot}$-10$^{5.6}$M$_{\odot}$, there was a transition from the triple systems ending their lifetimes in escape to ending in a merger. As the mass unit increased, the lifetimes of the systems decreased exponentially. This was drastic for large cases of 10$^9$M$_{\odot}$ and above where lifetimes were zero since there was instantaneous merging of the black holes. In studying the effect of distance with fixed mass unit of 10$^6$M$_{\odot}$, a strong negative correlation was found between the fraction of simulations ending in mergers and increasing distance unit. The average number of two-body  encounters increased as distance unit increased, however, at the 10$^3$pc distance unit there was a notable spike. The average lifetimes of the systems increased exponentially with increasing distance unit. A subsequent study is planned by the authors to analyse the effect of spin on triple systems in these Pythagorean configurations. Below we discuss the broader astrophysical implications of these results.

\color{black}

\section*{Acknowledgements}
Authors are thankful to Prof. Seppo Mikkola for providing his code and for advice on its usage. We are also grateful to our reviewers for such valuable feedback and suggestions that greatly improved the manuscript.

\section*{Data Availability}
The data underlying this paper are available in [Figshare Repository], at 
\url{https://doi.org/10.6084/m9.figshare.13194080.v2} 

Code by Prof. Mikkola is available at - \url{http://www.astro.utu.fi/mikkola/}

Additional code by authors can be found here - \url{https://doi.org/10.6084/m9.figshare.13194146.v1}




\bibliographystyle{mnras}
\bibliography{references} 

\clearpage




\bsp	
\label{lastpage}
\end{document}